\documentclass[12pt]{article}
\usepackage{eurosym}
\usepackage{amsmath}
\usepackage{amssymb}
\usepackage{amsfonts}
\usepackage{amsthm}
\usepackage{verbatim}

\usepackage{hyperref}

\usepackage[
backend=biber,
style=apa,
sorting=nyt
]{biblatex}
\DeclareSourcemap{
  \maps[datatype=bibtex]{
    \map{
      \step[fieldset=issn, null]
      \step[fieldset=doi, null]
      \step[fieldset=url, null]
      \step[fieldset=urldate, null]
    }
  }
}

\hypersetup{
    colorlinks=true,
    urlcolor=blue,
    linkcolor=blue,
    citecolor=blue
}

\usepackage{geometry}
\newtheorem{thm}{Theorem}[section]
\newtheorem{cor}{Corollary}[section]

\newtheorem{prop}{Proposition}[section]
\newtheorem{ex}{Example}[section]

\newtheorem{defn}{Definition}[section]

\newcommand{\comments}[1]{}
\numberwithin{equation}{section}

\newtheorem{ass}{Assumption}[section]
\input{tcilatex}

\begin{document}

\title{Entropy methods for identifying hedonic models}
\author{Arnaud Dupuy\thanks{%
E-mail: arnaud.dupuy@ceps.lu.} \\
CEPS/INSTEAD \and Alfred Galichon\thanks{%
Corresponding author. E-mail: alfred.galichon@sciencespo.fr. Galichon's
research has received funding from the European Research Council under the
European Union's Seventh Framework Programme (FP7/2007-2013) / ERC grant
agreement no 313699, and from FiME, Laboratoire de Finance des March\'{e}s
de l'Energie.} \\
Sciences Po \and Marc Henry\thanks{%
E-mail: marc.henry@psu.edu. Henry's research is supported by SSHRC Grant
435-2013-0292 and NSERC Grant 356491-2013.} \\
Penn State University}
\date{June 12, 2014\\
Dedicated to Ivar Ekeland on his 70th birthday\thanks{
This paper has benefited from insightful conversations with Ivar Ekeland and
Bernard Salani\'{e}. We would like to thank an anonymous referee for
comments on an earlier version of the paper.}.}
\maketitle

\begin{abstract}
This paper contributes to the literature on hedonic models in two ways.
First, it makes use of Queyranne's reformulation of a hedonic model in the
discrete case as a network flow problem in order to provide a proof of
existence and integrality of a hedonic equilibrium and efficient computation
of hedonic prices. Second, elaborating on entropic methods developed in
Galichon and Salani\'{e} (2014), this paper proposes a new identification
strategy for hedonic models in a single market. This methodology allows one
to introduce heterogeneities in both consumers' and producers' attributes
and to recover producers' profits and consumers' utilities based on the
observation of production and consumption patterns and the set of hedonic
prices.
\end{abstract}

\noindent

\section{Introduction\label{sec:Hedonics}}

Starting with Court (1941), Griliches (1961) and Lancaster (1966), a large
literature has aimed at providing a theoretical framework for pricing the
attributes of highly differentiated goods. While this literature was
initially mainly empirical in nature and early contributions lacked a proper
theoretical setting, the first theoretical treatments of hedonic models
appeared in Tinbergen (1956) and Rosen (1974). Tinbergen (1956) presents a
stylized model in which preferences are quadratic and attributes normally
distributed. Rosen (1974) showed the theoretical relation of hedonic prices
to marginal willingness to produce and marginal willingness to consume.
Hedonic models have also been used to study the pricing of highly
differentiated products such as houses (Kain and Quigley, 1970), wine and
champagne (Golan and Shalit, 1993), automobiles (Triplett, 1969) among
others, but also set forth a new literature on the Value of Statistical Life
following Thaler and Rosen's (1976) original idea of seeing jobs attributes
and in particular \textquotedblleft risk taken on the job\textquotedblright\
as a vector of hedonic attributes valued on the labor market. More recently,
significant progress on the understanding of the properties of hedonic
models (properties of an equilibrium, identification of deep parameters
etc.) has been achieved. These developments are to a large extent due to
Ivar Ekeland's contributions, see e.g. Ekeland et al. (2004) and Ekeland
(2010a, 2010b), and it is a pleasure to dedicate to him the present piece of
work in recognition of our intellectual debt to him.

In this paper we contribute to the hedonic literature in two ways. First, we
elaborate on an idea of Maurice Queyranne who reformulated the hedonic model
in the discrete case as a network flow problem. This reformulation allows us
to derive results on the existence of a hedonic equilibrium in the discrete
case, and it allows the use of powerful computational techniques to solve
for the equilibrium. Second, building on recent development in the matching
model literature and in particular the seminal contribution due to Choo and
Siow (2006) generalized by Galichon and Salani\'{e} (2014), we introduce
heterogeneities (unobserved by the econometrician) in producer and consumer
types. This formalism has two advantages: (i) it allows for the
incorporation of unobserved heterogeneity in the producers and consumers
characteristics, and (ii) it provides straightforward identification
results. Indeed, we follow Galichon and Salani\'{e} in making use of the
convex duality in discrete choice problems to recover utilities from choice
probabilities on both side of the market.

The remainder of the paper is organized as follows. Section~\ref%
{par:Hedonics-Discrete-Equil} discusses the properties of an equilibrium in
hedonic models and its reformulation as a network flow problem. Section~\ref%
{par:Hedonics-Discrete-Heterog} introduces a model with unobserved
heterogeneities on both sides of the market and studies the identification
of preference parameters. The discussion in Section~\ref{sec:Discussion}
concludes the paper.

\section{Equilibrium, existence and properties\label%
{par:Hedonics-Discrete-Equil}}

\subsection{Hedonic equilibrium}

\textbf{The model}. Throughout this paper, $\mathcal{X}$ is the set of
observable types of producers of a given good, and $\mathcal{Y}$ the set of
observable types of consumers of that good. This good comes in various
qualities; let $\mathcal{Z}$ be the set of the good's qualities. The sets $%
\mathcal{X}$, $\mathcal{Y}$ and $\mathcal{Z}$ are assumed to be finite. It
is assumed that there is a supply $n_{x}$ (resp. $m_{y}$) of producers
(resp. consumers) of type $x$ (resp. $y$). It is assumed that producers
(resp. consumers) can produce (consume) at most one unit of good. They have
the option not to participate in the market, in which case they choose $z=0$.

For example, hedonic models can be used to model the market for fine wines%
\footnote{%
We are confident Ivar will approve of this choice of example.}. In that
case, $\mathcal{X}$ may be the set of observable characteristics of wine
producers (say, grapes used, average amount of sunshine, and harvesting
technology), and $\mathcal{Y}$ may be the set of observable characteristics
of wine consumers (say country and purchasing channel). $\mathcal{Z}$ will
be the quality of the wine (say acidity, sugar content, expert rating).

\bigskip

Let $p_{z}$ be the price of the good of quality $z$. If a producer of type $%
x $ produces the good in quality $z$, the payoff to the producer is $\alpha
_{xz}+p_{z}$, where $\alpha _{xz}\in \mathbb{R}\cup \left\{ -\infty \right\} 
$ is the producer's productivity (the opposite of a production cost).
Similarly, if the consumer of type $y$ consumes the good in quality $z$, the
payoff to the consumer is $\gamma _{zy}-p_{z}$, where $\gamma _{zy}\in 
\mathbb{R}\cup \left\{ -\infty \right\} $ is the utility of the consumer%
\footnote{%
Note that in this setup, the utility of agents on each side of the market
does not depend directly on the type of the agent with whom they match, only
through the type of the contract. A more general framework where $\alpha $
and $\gamma $ depend simultaneously on $x$, $y$ and $z$ is investigated in
Dupuy, Galichon and Zhao (2014).}. Producers and consumers who do not
participate in the market get a surplus of zero.

\bigskip

\textbf{Supply and demand}. Let $\mu _{xz}$ be the supply function, that is
the number of producers of type $x$ offering quality $z$; similarly, $\mu
_{zy}$ is the demand function, the number of consumers of type $y$ demanding
quality $z$. One has%
\begin{equation*}
\sum_{z\in \mathcal{Z}}\mu _{xz}\leq n_{x}~,~\sum_{z\in \mathcal{Z}}\mu
_{zy}\leq m_{y}
\end{equation*}%
where the difference between the right-hand side and the left-hand side of
these inequalities is the number of producers of type $x$ (resp. consumers
of type $y$) deciding to opt out of the market. The market clearing
condition for quality $z$ expresses that the total quantity of good of
quality $z$ produced is equal to the total quantity consumed, that is%
\begin{equation*}
\sum_{x\in \mathcal{X}}\mu _{xz}=\sum_{y\in \mathcal{Y}}\mu _{zy}
\end{equation*}%
(it is assumed that there is no free disposal; if free disposal is assumed
the equality is replaced by $\geq $ in the expression).

\bigskip

\textbf{Equilibrium prices}. At equilibrium, each producer $x$ will optimize
its production behavior given the price vector $\left( p_{z}\right) $; hence
if producing quality $z^{\prime }$ yields strictly more profit than
producing quality $z$, then quality $z$ will not be produced at all; that is 
$\alpha _{xz}+p_{z}<\alpha _{xz^{\prime }}+p_{z^{\prime }}$ for some $%
z^{\prime }$ implies $\mu _{xz}=0$. A similar condition holds for consumers.

\bigskip

One can now state a formal definition.

\begin{defn}[Hedonic equilibrium]
\label{def:HedonEq}Let $\left( p_{z}\right) _{z\in \mathcal{Z}}$ be a price
vector, $\mu _{xz}$ a supply function, and $\mu _{zy}$ a demand function.
Then:

(a) $\left( p,\mu \right) $ is called a \emph{hedonic equilibrium} whenever
the following three conditions are all verified:

(i) People counting: the number of producers of type $x$ actually
participating in the market does not exceed the total number of agents of
type $x$, and similarly for consumers of type $y$. That is, for any $x$ and $%
y$, 
\begin{equation}
\sum_{z}\mu _{xz}\leq n_{x}~,~\sum_{z}\mu _{zy}\leq m_{y}.
\label{peopleCounting}
\end{equation}

(ii) Market clearing: for any $z$, supply for quality $z$ will equate
demand, that is%
\begin{equation}
\sum_{x\in \mathcal{X}}\mu _{xz}=\sum_{y\in \mathcal{Y}}\mu _{zy}.
\label{marketClearing}
\end{equation}

(iii) Rationality: no producer or consumer chooses a quality that is
sub-optimal. That is, given $\left( x,y,z,z^{\prime }\right) $, then%
\begin{eqnarray*}
\alpha _{xz}+p_{z} &<&\alpha _{xz^{\prime }}+p_{z^{\prime }}\text{ implies }%
\mu _{xz}=0 \\
\gamma _{zy}-p_{z} &<&\gamma _{z^{\prime }y}-p_{z^{\prime }}\text{ implies }%
\mu _{zy}=0.
\end{eqnarray*}

\bigskip

(b) If $n_{x}$ and $m_{y}$ are integer, $\left( p,\mu \right) $ is called an 
\emph{integral equilibrium} whenever $\left( p,\mu \right) $ is a hedonic
equilibrium and all the entries $\mu $ are integers.
\end{defn}

\bigskip

The indirect utility $u_{x}$ of a producer of type $x$ and the indirect
utility $v_{y}$ of a consumer of type $y$ are given by $u_{x}=\max_{z}\left(
\alpha _{xz}+p_{z},0\right) $ and $v_{y}=\max_{z}\left( \gamma
_{zy}-p_{z},0\right) $.

As a result, if $p_{z}$ is an equilibrium price, then for all $x$, $y$ and $%
z $, $u_{x}\geq \alpha _{xz}+p_{z}$ and $v_{y}\geq \gamma _{zy}-p_{z}$, thus 
$\gamma _{zy}-v_{y}\leq p_{z}\leq u_{x}-\alpha _{xz}$. Therefore:

\begin{prop}
For a given optimal solution $u$ and $v$, the set of equilibrium prices are
the prices $p_{z}$ such that%
\begin{equation}
p_{z}^{\max }\geq p_{z}\geq p_{z}^{\min }.  \label{RightPrice}
\end{equation}%
where 
\begin{equation}
p_{z}^{\min }=\max_{y}\left( \gamma _{zy}-v_{y}\right) \text{ and }%
p_{z}^{\max }=\min_{x}\left( u_{x}-\alpha _{xz}\right) .  \label{PriceBounds}
\end{equation}
\end{prop}

As a result, $u_{x}+v_{y}\geq \alpha _{xz}+\gamma _{zy}$, hence%
\begin{equation}
u_{x}+v_{y}\geq \max_{z}\left( \alpha _{xz}+\gamma _{zy}\right) ,
\label{eq:indirect}
\end{equation}%
thus, as observed by Chiappori, McCann and Nesheim (2010), $u$ and $v$ are
the stable payoffs of the assignment game in transferable utility with
surplus $\Phi _{xy}=\max_{z}\left( \alpha _{xz}+\gamma _{zy}\right) $. In
the next paragraph, we shall go beyond this equivalence through a
reformulation of the hedonic model as a network flow problem.

\subsection{Network flow formulation}

Interestingly, as understood by Maurice Queyranne, the hedonic equilibrium
problem can be reformulated as a network flow problem. This reformulation is
of particular interest since, as we show below, it helps us establish the
existence of a hedonic equilibrium and provides the building blocks to
compute an equilibrium. While the present exposition is as self-contained as
possible, a good reference for network flow problems is Ahuja, Magnanti and
Orlin (1993).

\bigskip

\textbf{The network}. Define a set of \emph{nodes} by $\mathcal{N}=\mathcal{X%
}\cup \mathcal{Z}\cup \mathcal{Y}$, and a set of \emph{arcs} $\mathcal{A}$
which is a subset of $\mathcal{N}\times \mathcal{N}$ and is such that if $%
ww^{\prime }\in \mathcal{A}$, then $w^{\prime }w\notin \mathcal{A}$. Here,
the set of arcs is $\mathcal{A}=\left( \mathcal{X}\times \mathcal{Z}\right)
\cup \left( \mathcal{Z}\times \mathcal{Y}\right) $.

\bigskip

A \emph{vector }is defined as an element of $\mathbb{R}^{\mathcal{A}}$.
Here, we introduce the following \emph{direct surplus vector} 
\begin{subequations}
\label{DefPhi}
\begin{eqnarray}
\phi _{ww^{\prime }} &:&=\alpha _{xz}\text{ if }w=x\text{ and }w^{\prime }=z
\\
\phi _{ww^{\prime }} &:&=\gamma _{zy}\text{ if }w=z\text{ and }w^{\prime }=y.
\end{eqnarray}

\bigskip

For two nodes $w$ and $w^{\prime }$, a \emph{path} from $w$ to $w^{\prime }$
is a chain 
\end{subequations}
\begin{equation*}
(w_{0}w_{1}),(w_{1}w_{2}),...,(w_{T-2}w_{T-1}),(w_{T-1}w_{T})
\end{equation*}%
such that $w_{i}w_{i+1}\in \mathcal{A}$ for each $i$, $w_{0}=w$ and $%
w_{T}=w^{\prime }$. $T$ is the \emph{length} of the path. Here, the only
nontrivial paths are of length 2 and are of the form $\left( xz\right)
,\left( zy\right) $ where $x\in \mathcal{X}$, $z\in \mathcal{Z}$ and $y\in 
\mathcal{Y}$.

\bigskip

For two nodes $w$ and $w^{\prime }$, we define the \emph{reduced surplus},
or \emph{indirect surplus} as the surplus associated with the optimal path
from $w$ to $w^{\prime }$. Here, for $x\in \mathcal{X}$, $y\in \mathcal{Y}$,
the indirect suplus $\Phi _{xy}$ of producer $x$ and consumer $y$ is 
\begin{equation}
\Phi _{xy}:=\max_{z\in \mathcal{Z}}\left( \alpha _{xz}+\gamma _{zy}\right) .
\label{FormuleChiappoMcCannNesheim}
\end{equation}

\bigskip

For $w\in \mathcal{N}$, we let $N_{w}$ be the algebraic quantity of mass
leaving the network at $w$. Hence $N_{w}$ is the flow of mass being consumed
($N_{w}>0$) or produced ($N_{w}<0$) at $w$. The nodes such that $N_{w}<0$
(resp. $N_{w}=0$ and $N_{w}>0$) are called the source nodes, whose set is
denoted $\mathcal{S}$ (resp. intermediate nodes $\mathcal{I}$ and target
nodes $\mathcal{T}$). Here, for $x\in \mathcal{X}$, $y\in \mathcal{Y}$, and $%
z\in \mathcal{Z}$, we set 
\begin{equation}
N_{x}:=-n_{x}~,~N_{y}:=m_{y}~,~N_{z}:=0  \label{DefN}
\end{equation}%
so that the set of source nodes is $\mathcal{S}:=\mathcal{X}$, the set of
intermediate nodes is $\mathcal{I}:=\mathcal{Z}$, and the set of target
nodes is $\mathcal{T}:=\mathcal{Y}$.

\bigskip

\textbf{Gradient, flow}. We define a \emph{potential} as an element of $%
\mathbb{R}^{\mathcal{N}}$. We define the \emph{gradient matrix} as the
matrix $\nabla $ of general term $\nabla _{aw}$, $a\in \mathcal{A}$, $w\in 
\mathcal{N}$ such that%
\begin{equation*}
\nabla _{aw}=-1\text{ if }a=ww^{\prime }\text{ for some }w^{\prime }\in 
\mathcal{N}\text{, }\nabla _{aw}=1\text{ if }a=w^{\prime }w\text{ for some }%
w^{\prime }\in \mathcal{N}\text{,}
\end{equation*}%
so that, for a potential $f\in \mathbb{R}^{\mathcal{N}}$, $\nabla f$ is the
vector such that for $a=ww^{\prime }\in \mathcal{A}$, one has $\left( \nabla
f\right) _{ww^{\prime }}=f_{w^{\prime }}-f_{w}$. Here, set the potential of
surpluses $U$ as 
\begin{equation}
U_{x}:=-u_{x}~,~U_{z}:=-p_{z}~,~U_{y}:=v_{y},  \label{DefU}
\end{equation}%
and 
\begin{equation}
\left( \nabla U\right) _{xz}=u_{x}-p_{z}\text{ and }\left( \nabla U\right)
_{zy}=v_{y}+p_{z}.  \label{EquivGradU}
\end{equation}

\bigskip

We define the \emph{divergence }matrix $\nabla ^{\ast }$ (sometimes also
called \emph{node-edge}, or \emph{incidence matrix}\footnote{%
The node-edge matrix is usually denoted $A$; our notations $\nabla ^{\ast }$
and terminology are chosen to stress the analogy with the corresponding
differential operators in the continuous case.}) as the transpose of the
gradient matrix: $\nabla _{xa}^{\ast }:=\nabla _{ax}$. As a result, for a
vector $v$, 
\begin{equation*}
\left( \nabla ^{\ast }v\right) _{ww^{\prime }}=\sum_{z}v_{zw^{\prime
}}-\sum_{z}v_{wz}.
\end{equation*}

A \emph{flow} is a nonnegative vector $\mu \in \mathbb{R}_{+}^{\mathcal{A}}$
that satisfies the \emph{balance of mass equation}\footnote{%
In most physical systems, mass is conserved and the balance equation has the
more usual form of \emph{Kirchoff's law} $\nabla ^{\ast }\mu =N$. However,
in the present setting, producers and consumers have an option not to
participate in the market, hence $\nabla ^{\ast }\mu =N$ is replaced by Eqs.
(\ref{k1})-(\ref{k3}).}, that is 
\begin{eqnarray}
\left( N-\nabla ^{\ast }\mu \right) _{w} &\geq &0,~w\in \mathcal{S}
\label{k1} \\
\left( N-\nabla ^{\ast }\mu \right) _{w} &=&0,~w\in \mathcal{I}  \label{k2}
\\
\left( N-\nabla ^{\ast }\mu \right) _{w} &\leq &0,~w\in \mathcal{T}
\label{k3}
\end{eqnarray}

Here, $\mu :\left( \mu _{xz},\mu _{zy}\right) $ is a flow if and only if $%
\mu _{xz}$ and $\mu _{zy}$ satisfy the people counting and market clearing
equations, that is 
\begin{equation*}
\sum_{z}\mu _{xz}\leq n_{x}~,~\sum_{z}\mu _{zy}\leq m_{y}\text{ and }%
\sum_{x\in \mathcal{X}}\mu _{xz}=\sum_{y\in \mathcal{Y}}\mu _{zy}.
\end{equation*}

\bigskip

\textbf{Maximum surplus flow}. We now consider the $\emph{maximum\ surplus\
flow\ problem}$, that is 
\begin{eqnarray}
&&\max_{\mu \in \mathbb{R}_{+}^{\mathcal{A}}}\sum_{a\in \emph{A}}\mu
_{a}\phi _{a}  \label{netFlow} \\
s.t.~ &&\mu \text{ satisfies (\ref{k1}), (\ref{k2}), (\ref{k3}),}  \notag
\end{eqnarray}%
whose value coincides with the value of its dual version, that is%
\begin{eqnarray}
&&\min_{U\in \mathbb{R}^{\mathcal{N}}}\sum_{w\in \mathcal{N}}U_{w}N_{w}
\label{dualNetflow} \\
s.t. &&U_{w}\geq 0,~\forall w\in \mathcal{S}\cup \mathcal{T}  \notag \\
&&\nabla U\geq \phi ,  \notag
\end{eqnarray}%
and by complementary slackness, for $w\in \mathcal{S}\cup \mathcal{T},$ $%
U_{w}>0$ implies $N_{w}=\left( \nabla ^{\ast }\mu \right) _{w}$. A standard
result is that if $N$ has only integral entries, then (\ref{netFlow}) has an
integral solution $\mu $.

\bigskip

Here the solution $U$ of (\ref{dualNetflow}) is related to the solution to
the hedonic model by Equations (\ref{DefU}), that is $%
u_{x}=-U_{x}~,~p_{z}=-U_{z}~,~v_{y}=U_{y}$. Using (\ref{EquivGradU}) and (%
\ref{DefPhi}), $\nabla U\geq \phi $ implies $u_{x}-p_{z}=U_{z}-U_{x}\geq
\phi _{xz}=\alpha _{xz}$ and $v_{y}+p_{z}=U_{y}-U_{z}\geq \phi _{zy}=\gamma
_{zy}$, thus, using complementary slackness one recovers 
\begin{equation*}
u_{x}=\max_{z}\left( \alpha _{xz}+p_{z}\right) ^{+}\text{ and }%
v_{y}=\max_{z}\left( \gamma _{zy}-p_{z}\right) ^{+}.
\end{equation*}%
Further, if $n$ and $m$ have only integral entries, then there is an
integral solution $\mu $ to (\ref{netFlow}). Therefore:

\begin{thm}[Queyranne]
\label{thm:HedonicsAsMatchingFlow}The hedonic equilibrium problem of
Definition~\ref{def:HedonEq} can be reformulated as a maximum surplus flow
problem as described above.
\end{thm}

\bigskip

As announced above, this reformulation has several advantages. First, it
establishes the existence of a hedonic equilibrium, and its integrality.

\begin{thm}[Existence]
\label{thm:hedoneq-exist}Consider a market given by $n_{x}$ producers of
type $x$, $m_{y}$ consumers of type $y$, and where productivity of producer $%
x$ is given by $\alpha _{xz}$, and utility of consumer $y$ is $\gamma _{zy}$%
.Then:

(i) There exists a hedonic equilibrium $\left( p_{z},\mu _{xz},\mu
_{zy}\right) $;

(ii) $\left( \mu _{xz},\mu _{zy}\right) $ are solution to the primal problem
of the expression of the social welfare 
\begin{eqnarray}
&&\max_{\mu _{xz},\mu _{zy}\geq 0}\sum_{xz}\mu _{xz}\alpha
_{xz}+\sum_{yz}\mu _{zy}\gamma _{zy}  \label{HedonPrimal} \\
\sum_{z}\mu _{xz} &\leq &n_{x}\text{ and }\sum_{z}\mu _{zy}\leq m_{y}\text{
and }\sum_{x}\mu _{xz}=\sum_{y}\mu _{zy},  \notag
\end{eqnarray}%
while $\left( p_{z}\right) $ is obtained from the solution of the dual
expression of the social welfare%
\begin{eqnarray}
&&\min_{u_{x},v_{y}\geq 0;p_{z}}\sum_{x}n_{x}u_{x}+\sum_{y}m_{y}v_{y}
\label{HedonDual} \\
u_{x} &\geq &\alpha _{xz}+p_{z}\text{ and }v_{y}\geq \gamma _{zy}-p_{z}. 
\notag
\end{eqnarray}%
expressed equivalently as 
\begin{equation*}
\min_{p_{z}}\{\sum_{x}n_{x}\max_{z}\left( \alpha _{xz}+p_{z},0\right)
+\sum_{y}m_{y}\max_{z}\left( \gamma _{zy}-p_{z},0\right) \}.
\end{equation*}

(iii) If $n_{x}$ and $m_{y}$ are integral for each $x$ and $y$, then $\mu
_{xz}$ and $\mu _{zy}$ can be taken integral.
\end{thm}

\bigskip

Second, on the practical side, Theorem \ref{thm:hedoneq-exist} also has a
useful consequence in terms of computation of the equilibrium, as shown in
the following corollary.

\begin{cor}
The equilibrium prices $\left( p_{z}\right) $ as well as the quantities $\mu
_{xz},\mu _{zy}$ supplied at equilibrium can be determined using one of the
many minimum cost flow algorithms, see for instance Ahuja, Magnanti and
Orlin (1993).
\end{cor}

\begin{ex}
Assume that there are four sellers and three buyers, each of whom is unique
among her type, and three qualities. Participation is endogenous but there
is no free disposal. Assume that the technology and preference parameters
are given by 
\begin{equation*}
\left( \alpha _{xz}\right) =\left( 
\begin{array}{ccc}
2 & 5 & 3 \\ 
2 & 1 & 4 \\ 
1 & 5 & 8 \\ 
4 & 2 & 4%
\end{array}%
\right) \text{ and }\left( \gamma _{zy}\right) =\left( 
\begin{array}{ccc}
0 & 2 & 1 \\ 
2 & 4 & 2 \\ 
4 & 2 & 6%
\end{array}%
\right) .
\end{equation*}

The indirect utilities of the buyers and the sellers are determined by
linear programming. One finds $u_{x}^{\min }=\left( 0~0~4~0\right) $ and $%
v_{y}^{\max }=\left( 8~9~10\right) $, and $u_{x}^{\max }=\left(
3~0~4~0\right) $, and $v_{y}^{\min }=\left( 8~6~10\right) $, and the optimal
matching will consist in matching $x_{1}$ with $y_{2}$, which produce
together quality 2, and any other two remaining producers with the two other
remaining consumers, producing two units of quality of quality 3. Hence the
optimal number of goods produced in each quality, denoted $l$, is given by $%
l_{x_{1}}=0$, $l_{x_{2}}=1$ and $l_{x_{3}}=2$. Making use of $p_{z}^{\min
}=\max_{y}\left( \gamma _{zy}-v_{y}^{\max }\right) $ and $p_{z}^{\max
}=\min_{x}\left( u_{x}^{\max }-\alpha _{xz}\right) ,$one finds that if $%
u=\left( 0~0~4~0\right) $ and $v=\left( 8~9~10\right) $, then $p\in \left[
-7,-4\right] \times \left[ -5,-2\right] \times \left\{ -4\right\} $.
\end{ex}

\section{Introducing heterogeneities\label{par:Hedonics-Discrete-Heterog}}

In the spirit of Galichon and Salani\'{e} (2014), who extended the model of
Choo and Siow (2006), we are now going to introduce heterogeneities in
producers' and consumers' characteristics. As before, we consider the set $%
\mathcal{X}$ of observable types of producers, the set $\mathcal{Y}$ of
observable types of consumers, and the set $\mathcal{Z}$ of qualities, and
the sets $\mathcal{X}$, $\mathcal{Y}$ and $\mathcal{Z}$ are finite\footnote{%
However, the ideas presented here extend to the continuous case, see Dupuy
and Galichon (2014) for a continuous logit approach and Chernozhukov,
Galichon and Henry (2014)\ for an approach based on multivariate quantile
maps.}. In the sequel, $i$ will denote an individual producer, and $j$ will
denote an individual consumer. The analyst observes the \textquotedblleft
observable type\textquotedblright\ $x_{i}\in \mathcal{X}$ of producer $i$,
and the \textquotedblleft observable type\textquotedblright\ $y_{j}\in 
\mathcal{Y}$ of consumer $j$. Two producers (resp. consumers) sharing the
same observable type may differ in some additional heterogeneity term that
will affect their profitability (resp. utility) function. This heterogeneity
is observed by the consumers but not by the analyst. It is assumed that the
quality $z\in \mathcal{Z}$ is fully observable by all parties and the
analyst.

\bigskip

If the price of quality $z$ is $p_{z}$, then the profit of an individual
producer $i$ selling quality $z$ is defined as $\tilde{\alpha}_{iz}+p_{z}\in 
\mathbb{R}\cup \left\{ -\infty \right\} $, and the utility of an individual
consumer $j$ purchasing $z$ is defined as $\tilde{\gamma}_{jz}-p_{z}\in 
\mathbb{R}\cup \left\{ -\infty \right\} $. If producer $i$ (resp. consumer $%
j $) does not participate in the market, she gets a surplus of $\tilde{\alpha%
}_{i0}$ (resp. $\tilde{\gamma}_{j0}$). The tilde notation in $\tilde{\alpha}$
and $\tilde{\gamma}$ indicates that these terms characterize the invididual
level, which will be random from the point of view of the observer. Note
that the utility of agents on each side of the market still does not depend
directly on the type of the agent with whom they match, but only indirectly
via the type of the contract.

\bigskip

\subsection{Structure of the heterogeneity}

We introduce an structural assumption regarding the structure of unobserved
heterogeneity.

\begin{ass}
Assume that the pre-transfer profitability and utility terms have structure 
\begin{eqnarray*}
\tilde{\alpha}_{iz} &=&\alpha _{x_{i}z}+\varepsilon _{iz}\text{ and }\tilde{%
\gamma}_{jz}=\gamma _{y_{j}z}+\eta _{jz} \\
\tilde{\alpha}_{i0} &=&\varepsilon _{i0}\text{ and }\tilde{\gamma}_{j0}=\eta
_{j0}
\end{eqnarray*}%
where:

a) The surplus shock, or unobserved heterogeneity component $\varepsilon
_{i} $ of all producers of a given observable characteristics $x$ are drawn
from the same distribution $\mathbf{P}_{x}$.

b) The surplus shock, or unobserved heterogeneity component $\eta _{j}$ of
all consumers of a given observable characteristics $y$ are drawn from the
same distribution $\mathbf{Q}_{y}$.

c) The distributions $\mathbf{P}_{x}$ and $\mathbf{Q}_{y}$ have full support.
\end{ass}

Part a) and b) of this assumption are not very restrictive. They essentially
express that the quality $z$ is fully observed. Part c) is more restrictive.
It implies that for each type of producer or consumer, and for any quality,
some individual of this type will produce or consume this quality. This
assumption does not hold if, say, some technological constraint prevents
some producers from producing a given quality. Although this assumption is
not required, and is not needed in Galichon and Salani\'{e} (2014), it
greatly simplifies the results on identification and we will maintain it for
the purposes of this paper.

\bigskip

We will also assume that:

\begin{ass}
There is a large number of producers and consumers of each given observable
type, and each of them are price takers.
\end{ass}

This assumption has two virtues. First, it implies that we can have a
statistical description of the producers and the consumer of a given type
and we do not need to worry about sample variation. Second, it rules out any
strategic behaviour by agents: the market here is assumed perfectly
competitive.

\bigskip

\subsection{Social welfare}

We now investigate the social welfare, understood as the sum of the
producers' and consumers' surpluses. We first focus on the side of
producers. At equilibrium, producer $i$ will get utility%
\begin{equation*}
U_{x_{i}z}+\varepsilon _{iz}
\end{equation*}%
from producing quality $z$, where 
\begin{equation*}
U_{xz}=\alpha _{xz}+p_{z}.
\end{equation*}

The sum of the ex-ante indirect surpluses of the producers of observable
type $x$ is $n_{x}G_{x}(U_{x\cdot })$, where $G_{x}(U_{x\cdot })$ is the
expected indirect utility of a consumer of type $x$, that is%
\begin{equation}
G_{x}(U_{x\cdot })=\mathbb{E}_{\mathbf{P}_{x}}\left[ \max_{z\in \mathcal{Z}%
}(U_{xz}+\varepsilon _{iz},\varepsilon _{i0})|x_{i}=x\right]
\label{ExpectedUtilityDC}
\end{equation}%
where the argument of $G_{x}$ is the $\left\vert \mathcal{Z}\right\vert $%
-dimensional vector of $\left( U_{xz}\right) _{z\in \mathcal{Z}}$, which is
denoted $U_{x\cdot }$, and where the expectation is taken with respect to
the distribution $\mathbf{P}_{x}$ of unobserved heterogeneity component $%
\varepsilon _{i}$. We refer to Galichon and Salani\'{e} (2014) for
mathematical properties of $G$ and examples. By the Envelope theorem, the
number of producers of type $x$ choosing quality $z$, denoted $\mu _{z|x}$,
is given by%
\begin{eqnarray}
\mu _{z|x}=\frac{\mu _{xz}}{n_{x}}= &&\mathbf{P}_{x}\left( x\text{ chooses }%
z\right)  \notag \\
&=&\frac{\partial G_{x}(U_{x\cdot })}{\partial U_{xz}}.  \label{EnvelopeThm}
\end{eqnarray}

\bigskip

This result sheds light on the \emph{equilibrium characterization problem}:
based on the vector of producer surpluses $U$, this allows to deduce the
production patterns $\mu $, and a similar picture holds on the consumers'
side. However, the \emph{identification problem} consists in recovering
utility parameters, here $U_{x\cdot }$ based on the observation of producer'
choices, here summarized by $\mu _{xz}$, the number of producers of
observable type $x$ who choose to sell quality $z$. This requires inverting
relation (\ref{EnvelopeThm}). To do this, still following Galichon and Salani%
\'{e} (2014), introduce the Legendre-Fenchel transform $G_{x}^{\ast }$ of $%
G_{x}$ as%
\begin{eqnarray}
G_{x}^{\ast }(\mu _{\cdot |x}) &=&\max_{U_{xz}}\left( \sum_{z\in \mathcal{Z}%
}\mu _{z|x}U_{xz}-G_{x}(U_{x\cdot })\right) \text{ if }\sum_{z\in \mathcal{Z}%
}\mu _{z|x}\leq 1  \label{hedonics-conjug} \\
&=&+\infty \text{ otherwise.}  \notag
\end{eqnarray}%
where $\mu _{\cdot |x}$ is the vector of choice probabilities $\left( \mu
_{z|x}\right) _{z\in \mathcal{Z}}$. By the Envelope theorem, one has%
\begin{equation}
U_{xz}=\frac{\partial G_{x}^{\ast }(\mu _{\cdot |x})}{\partial \mu _{z|x}}.
\label{FOC}
\end{equation}

Hence $U_{xz}$ is identified from $\mu _{x.}$by equation (\ref{FOC}).
Galichon and Salani\'{e} (2014) have shown that $G^{\ast }$ can be very
efficiently computed as the solution to an optimal matching problem.

\bigskip

Similarly to the producers' side of the market, denote $V_{zy}=\gamma
_{zy}-p_{z}$ the deterministic part of the consumer's payoff from buying
good quality $z$, and write $V_{\cdot y}$ for the $|\mathcal{Z}|$%
-dimensional vector with $z$-th component $V_{zy}$. The sum of expected
utilities of consumers with observable characteristics $y$ is given by $%
m_{y}H_{y}(V_{\cdot y})$, where $H_{y}(V_{\cdot y})$ is the expected
indirect utility of a consumer of type $y$, that is%
\begin{equation*}
H_{y}(V_{\cdot y})=\mathbb{E}_{\mathbf{Q}_{y}}\left[ \max_{z\in \mathcal{Z}%
}(V_{zy}+\eta _{jz},\eta _{j0})|y_{j}=y\right] ,
\end{equation*}%
and $\mathbf{Q}_{y}$ is the distribution of the unobserved heterogeneity
component $\eta _{j}$ for a consumer indexed by $j$, with observable
characteristics $y=y_{j}$. Hence, as in the producer's case, we obtain
identification of $V_{zy}$ through the following relation. 
\begin{equation}
V_{zy}=\frac{\partial H_{y}^{\ast }(\mu _{\cdot |y})}{\partial \mu _{z|y}},
\label{FOCy}
\end{equation}%
where $H_{y}^{\ast }$ is the convex conjugate of $H_{y}$, defined by a
formula similar to (\ref{hedonics-conjug}).

\bigskip

Recall that the social welfare $\mathcal{W}$ is the sum of the producers and
consumers surpluses. We are now able to state the following result.

\begin{thm}
\label{theorem:SWF}(i) The optimal social welfare in this economy is given by%
\begin{equation}
\mathcal{W}=\min_{p}\sum_{x\in \mathcal{X}}n_{x}G_{x}\left( \alpha _{x\cdot
}+p_{\cdot }\right) +\sum_{y\in \mathcal{Y}}m_{y}H_{y}\left( \gamma _{\cdot
y}-p_{\cdot }\right) .  \label{SWFPrix}
\end{equation}

(ii) Alternatively, $\mathcal{W}$ can be expressed as%
\begin{eqnarray}
\mathcal{W} &=&\max_{\mu \geq 0}\sum_{x\in \mathcal{X},z\in \mathcal{Z}}\mu
_{xz}\alpha _{xz}+\sum_{y\in \mathcal{Y},z\in \mathcal{Z}}\mu _{zy}\gamma
_{zy}-\mathcal{E}\left( \mu \right)  \label{SWFDemand} \\
s.t.~ &&\mu \text{ satisfies (\ref{peopleCounting}) and (\ref{marketClearing}%
),}  \notag
\end{eqnarray}%
where $\mathcal{E}(\mu )$ is a generalized entropy function, defined by%
\begin{equation*}
\mathcal{E}(\mu )=\sum_{x\in \mathcal{X}}n_{x}G_{x}^{\ast }\left( \mu
_{x.}\right) +\sum_{y\in \mathcal{Y}}m_{y}H_{y}^{\ast }\left( \mu _{\cdot
y}\right) .
\end{equation*}

(iii) Further the equilibrium $\left( p_{z},\mu _{xz},\mu _{zy}\right) $ is
unique and is such that $\left( p_{z}\right) $ is a minimizer for (\ref%
{SWFPrix}) and $\left( \mu _{xz},\mu _{zy}\right) $ is a maximizer for (\ref%
{SWFDemand}).
\end{thm}

The terminology \textquotedblleft generalized entropy\textquotedblright\
comes from the fact, that in the Logit case where the utility shocks $%
\varepsilon $ and $\eta $ are i.i.d. and have a Gumbel distribution, then $%
\mathcal{E}(\mu )$ is a regular entropy function, namely%
\begin{equation*}
\mathcal{E}(\mu )=\sum_{x\in \mathcal{X},~y\in \mathcal{Y}}\mu _{xy}\log 
\frac{\mu _{xy}^{2}}{n_{x}m_{y}}+\sum_{x\in \mathcal{X}}\mu _{xy}\log \frac{%
\mu _{x0}}{n_{x}}+\sum_{y\in \mathcal{Y}}\mu _{xy}\log \frac{\mu _{0y}}{m_{y}%
}
\end{equation*}%
where $\mu _{x0}=n_{x}-\sum_{z\in \mathcal{Y}}\mu _{xz}$ and $\mu
_{0y}=m_{y}-\sum_{z\in \mathcal{Y}}\mu _{zy}$.

\bigskip

\subsection{Identification}

As a result of the first order conditions in the previous theorem, the model
is exactly identified from the observation of the hedonic prices $p_{z}$,
along with the production and consumption patterns $\mu _{xz}$ and $\mu
_{zy} $.

\begin{thm}
\label{theorem:identification}The producers and consumers systematic
surpluses at equilibrium $U$ and $V$ are identified from $\mu _{xz}$ and $%
\mu _{zy}$ by%
\begin{equation*}
U_{xz}=\frac{\partial G_{x}^{\ast }(\mu _{\cdot |x})}{\partial \mu _{z|x}}%
\text{ and }V_{zy}=\frac{\partial H_{y}^{\ast }(\mu _{\cdot |y})}{\partial
\mu _{z|y}}.
\end{equation*}

Hence $\alpha $ and $\gamma $ are identified from $\mu _{xz}$, $\mu _{zy}$
and $p_{z}$ by 
\begin{equation*}
\alpha _{xz}=\frac{\partial G_{x}^{\ast }(\mu _{\cdot |x})}{\partial \mu
_{z|x}}-p_{z}\text{ and }\gamma _{zy}=\frac{\partial H_{y}^{\ast }(\mu
_{\cdot |y})}{\partial \mu _{z|y}}+p_{z}.
\end{equation*}
\end{thm}

In the Logit case, these formulas become $\alpha _{xz}=\log (\mu _{xz}/\mu
_{x0})-p_{z}$ and $\gamma _{zy}=\log (\mu _{zy}/\mu _{0y})+p_{z}$, where $%
\mu _{x0}$ and $\mu _{0y}$ have been defined at the previous paragraph.

\bigskip

Note that (as it frequently is found in various situations in the
econometrics literature), the introduction of heterogeneity has allowed to
identify simultaneously $\alpha _{xz}$ and $\gamma _{zy}$. When there is no
heterogeneity, it is well known that simultaneous identification of these
parameters is not possible. This is due to the fact that, in the absence of
heterogeneity, the solution $\mu $ of the problem is no longer an interior
point, thus many entries $\mu _{xz}$ and $\mu _{zy}$ are forced to be equal
to zero.

\bigskip

\section{Discussion\label{sec:Discussion}}

The results presented in this paper are applicable to many different
empirical settings. Returning to the market for fine wines for example, the
analyst will typically have access to data about the share of consumers with
observable characteristics $y$ purchasing wine of quality $z$ and the share
of producers of type $x$ selling wine of quality $z$. Our methodology allows
to identify the surpluses of consumers and producers from these data. If in
addition, the price of wine of various qualities are observed, then the
utility $\alpha $\ of consumers and technology $\gamma $ of producers are
identified as well.

Next, consider the marriage market example. In classical models of sorting
on the marriage market, following Becker (1973) and Shapley and Shubik
(1972), the matching surplus between a man of type $x$ and a woman of type $%
y $ is 
\begin{equation*}
\Phi _{xy}=\alpha _{xy}+\gamma _{xy}
\end{equation*}%
where $\alpha $ and $\gamma $ are the man and the woman's surplus for being
married to each other. However, this analysis misses the fact that the
partners in the marriage market also need to make a number of joint
decisions, such as whether/when/how to raise children, where to live, how to
spend their spare time together, etc. This has the flavour of a hedonic
model. For the sake of discussion, consider (on the other extreme) a
framework where the observed characteristics is, say, the date of birth of
each agent, and where the only variable agents care about is, say, the date
of birth of their first child. In this context, the matching surplus is now%
\begin{equation*}
\Phi _{xy}=\sup_{z}(\alpha _{xz}+\gamma _{zy})
\end{equation*}%
and the methodology developed in this paper can identify the surplus of a
man born in $x=1985$ to have his first child in say $z=2012$ and the surplus
of a woman born in $y=1986$ to have her first child in $z=2013$. The
required data are the shares of men and women born in a given year who had
their first child in a given year. This example, however, is peculiar as men
and women are likely to form preferences not only over the hedonic attribute 
$z$, i.e. the year of birth of first child, but also over their spouse's
attributes $x$ and $y$. One therefore needs to consider a model encompassing
the hedonic model a la Rosen (1974) with the sorting model \`{a} la Becker
(1973). In this model, developed and studied in Dupuy and Galichon and Zhao
(2014) who apply it to the study of migration in China, the matching surplus
is%
\begin{equation*}
\Phi _{xy}=\sup_{z}(\alpha _{xzy}+\gamma _{xzy})
\end{equation*}%
and this model embeds both the classical sorting model ($\alpha
_{xzy}=\alpha _{xy}$ and $\gamma _{xzy}=\gamma _{xy}$) and the hedonic model
($\alpha _{xzy}=\alpha _{xz}$ and $\gamma _{xzy}=\gamma _{zy}$). The
empirically interesting question there is to assess which of the
\textquotedblleft sorting effect\textquotedblright\ or \textquotedblleft
hedonic effect\textquotedblright\ is strongest.

\end{document}